# Theoretical study of current density for InN and GaN-based Gunn diode operating in transit-time mode

**M O Islam**[1], **M G Hassan**[2], **M R Islam**[1], **M A Uddin**[1] **and Z H Mahmood**[1]
[1]Department of Applied Physics, Electronics and Communication Engineering, University of Dhaka, Dhaka-1000, Bangladesh
[2]Department of Mathematics, University of Dhaka, Dhaka-1000, Bangladesh

E-mail: m.onirbanislam@gmail.com

**Abstract.** We are the first to report the frequency response and corresponding current density of a wurtzite phase Indium Nitride (InN)-based vertical configuration Gunn diode at 1 µm active length. Domain growths dynamics with respect to space and time, variation of domain velocity and the field outside the vicinity of the domain with respect to domain field have also been studied using the presented mathematical equations. It has been found that, InN-based Gunn diodes are capable to operate around 224 GHz with 691 kA/cm$^2$ current density at 300 K. For comparison purpose, all these characteristics have been evaluated for Gunn diodes of different active lengths based on wurtzite phase Gallium Nitride (GaN). Simulated results are consistent with the other reports on GaN-based Gunn diode.

## 1. Introduction

Gunn oscillation is used as one of the basis for high frequency generation since its realization by Gunn in 1963. The theoretical framework was earlier proposed by Ridley and Watkins, Hilsum, and later developed by Kroemer [1]; Knight and Peterson [2]. Numerical studies have shown that Gunn oscillation can reach in the terahertz (THz) regime in nanoscale devices [3]. THz regime has found promising applications in medical, biological and industrial imaging, broadband and safety communication, radar, space science [4], automotive industry [5], or biomolecular hydration dynamics [6]. Negative differential resistance (NDR) is the key to enable Gunn oscillation in bulk semiconductors. It has been demonstrated by Monte Carlo simulation that the wurtzite phase binaries and ternaries of III-Nitride exhibit NDR in their velocity-field characteristics [7-9].

III-Nitrides, especially Gallium nitride (GaN) has received intensive attention due to its impressive performance in heterostructure field effect transistors [10, 11], ultraviolet light emitting diodes, optical switches, optical modulators, vertical-cavity lasers, nanowire lasers [12]. Existing reports [4, 13-19] have shown that GaN is capable to exhibit superior frequency and power handling performance than widely used GaAs based Gunn diode. Monte Carlo simulations have demonstrated an NDR relaxation frequency of 740 GHz in GaN while in GaAs it is only 109 GHz [4]. The NDR can arise from either the transferred-electron effect or as a consequence of the negative effective mass beyond the inflection point depending on the energy of the upper valley [20]. Experimentally, the first stable NDR is reported in 2007 by Yilmazoglu *et al* on vertical GaN-based Gunn device [21, 22].

Indium Nitride (InN) has been the least studied among III-nitrides. With the theoretical and experimental [12, 23] confirmation of narrow band gap along with very small effective mass and energy relaxation time, high electron mobility, peak overshoot velocity, and saturation velocity [12], InN is emerged as a promising material for high frequency electronic devices.

To evaluate the Gunn oscillation theoretically, it is necessary to incorporate their high-field transport characteristics in analytical models, or the use of some sort of simulations like Monte Carlo. It is our interest to evaluate the transit-time and current density of a Gunn diode operating in the transit-time mode. Analytical expressions for current density and transit time have been presented by Roblin and Rohdin [24]. They have described the excess field due to domain contribution in terms of the non-equilibrium carrier concentration qualitatively. The domain drift velocity is considered as uniform in computing the transit-time frequency. But the domain does not drift in a uniform motion until it reaches the saturation regime. Further, the concept of saturation velocity is not well defined for III-Vs [25]. Therefore, it is necessary to have an improved model to deduce different aspects of Gunn oscillation quantitatively. In the present report, an analytical model is presented considering the non uniform domain drift and the most recent reports on high-field transport for GaN [7, 8] and InN [12] unlike the previous reports [15 -19]. Three dimensional (3D) space and time evaluation of domain field, domain field width and the variation of the domain velocity have also been studied in order to enlighten the domain formation on the material basis. Finally, present work is the first theoretical report on the InN based Gunn diode.

## 2. Theoretical model

The assumptions have been taken in our model are: (I) electrons are the only charge carrier, (II) all the donors are ionized in the operating temperature of 300K and (III) one dimensional analysis has been carried out.

*2.1 Domain dynamics*

The fluctuation, in the carrier concentration has been considered due to the crystal defect, or doping inhomogenuities, or noise. The fluctuation can form a stable domain in the bulk semiconductor, if the device is biased in its NDR region [26]. The fluctuation in the carrier concentration about the equilibrium is

$$\delta n(x,t) = n(x,t) - n_0 \tag{1}$$

where, $n = n(x,t)$ and $n_0$ are the non-equilibrium and equilibrium carrier concentration, respectively; $x$ and $t$ are the space and time variable, respectively. Since small anisotropy has been observed in the velocity-field characteristics of GaN and InN [7, 8], the electric field due to the domain, $\boldsymbol{E}_{dom}$ is given by the Maxwell's equation:

$$\nabla_x \cdot \boldsymbol{E}_{dom} = -\frac{qn}{\varepsilon_0 \varepsilon_s}. \tag{2}$$

$\varepsilon_0$ is the permittivity of vacuum, $\varepsilon_s$, static dielectric constant of the bulk semiconductor, and q is the electronic charge. To find the space and time dependence of the carrier concentration $n(x,t)$, the current continuity equation, neglecting all kinds of generation and recombination events

$$\nabla_x \cdot \boldsymbol{J}_{dom} = -q\dot{n} \tag{3}$$

have to be coupled with Eq. (2); yielding [26]

$$\delta n(x) = \delta n(0) e^{-x/L_D} \tag{4a}$$

and

$$\delta n(t) = \delta n(0) e^{-t/\tau_D} \tag{4b}$$

Therefore, the fluctuation becomes

$$\delta n(x,t) = \delta n(x=0) e^{-\frac{x}{L_D}} \delta n(t=0) e^{-\frac{t}{\tau_D}} = \Delta n \, e^{-\frac{x}{L_D}} e^{-\frac{t}{\tau_D}} \tag{5}$$

where $J_{dom} = J_{dom}(x,t)$ is the domain current density and it has been assumed that $\delta n\,(x=0) = \delta n\,(t=0) = \delta n$ and $\delta n^2 = \Delta n$, a constant. The Debye length $L_D$ and dielectric differential relaxation time $\tau_D$ are defined in [1]. Assuming $n_0 = n(0,0) = N_D$, $N_D$ being the donor concentration, the carrier concentration in the NDR region can be given by from Eq. (1)

$$n(x,t) = N_D + \Delta n\, e^{-\frac{x}{L_D}} e^{-\frac{t}{\tau_D}} \tag{6}$$

The electric field ($E$) dependence of average drift velocity ($v_d$) for III-nitrides is given by [9, 12]

$$v_d(E) = \frac{\mu_0 E + v_{sat}\left(\frac{E}{E_{cr}}\right)^{\beta_1}}{1 + \left(\frac{E}{E_{cr}}\right)^{\beta_1} + a\left(\frac{E}{E_{cr}}\right)^{\beta_2}} \tag{7}$$

where $\mu_0 = \mu\,(T, N)$ is the electron temperature and doping dependent mobility, $v_{sat}$ is the saturatin velocity, $E_{cr}$ is the critical field, and $\alpha, \beta_1$, and $\beta_2$ are the fitting parameters.

*2.2 Current density*
When the Gunn diode is biased, i. e., $V_{DC} > 0V$, due to the doping inhomogenuities, or defects, or noise, there must be some fluctuation in the established electric field around these. The small fluctuation in the field eventually grows up, if the bias ensures that the device is in its NDR region. This perturbated region is termed as the domain field $E_{dom}$. At the moment the domain is formed, the field outside the domain, $E_r$ becomes smaller than the applied field. Eventually, there establishes two types of electric fields: $E_{dom}$ and $E_r$ [26]. Now from the equal-area rule [1] it is clear that

$$E_r = E_r(E_{dom})$$

and previously, we mentioned

$$E_{dom} = E_{dom}(x,t)$$

So, $E_r$ can be considered as a function of both space and time. But this is indeed not the situation when we want to calculate the current density due to the domain field. Because $E_{dom}$ is entirely a localized field while $E_r$ is a globalized field, as far as the diode physics is concerned. $E_r$ is uniformly distributed over the diode length except the domain vicinity [1]. Taking this aspect in account, we have

$$E_r = E_r(t)$$

Now the current density due to only $E_r$ can be given by

$$J_r(t) = qn_0 v_r(E_r) \tag{8}$$

Integrating over the whole active length L yields

$$\int_0^L J_r(t).\,dx = \int_0^L qn_0 v_r(E_r).\,dx \tag{9}$$

To get the contribution of the domain we proceed as follow. The domain will contribute in the external circuit when it touches the anode terminal, i.e., at $t = t_{tr}$, $t_{tr}$ be the transit time, mathematically the fact can be expressed as

$$J_{dom}(x,t) = 0 \qquad \forall\,(x,t) \neq (L, t_{tr}) \tag{10}$$

The current density $J_{dom}$ due to the domain, the conservation of charge principle can be used. Therefore, $J_{dom}$ can be calculated from

$$\nabla_x \cdot \boldsymbol{J}_{dom}]_{x=L} = -q\dot{n}]_{t=t_{tr}} \tag{11}$$

The domain field abruptly drops at the anode end, so as the carrier concentration responsible for the domain formation. Hence, $n$ is not a continuous function in the interval $[t_{tr}, t_{tr} + \Delta t]$, where $\Delta t \to 0$, hence the derivative of the right-hand side of Eq. (16) does not exists. To solve the problem, we consider $n$ as a multiple of the Heaviside step-function θ:

$$n(L, t_{tr}) = m\theta(-(t - t_{tr})) \tag{12}$$

where $m$ is a multiplier, represents the magnitude of $n(\xi, \xi)$. Considering only the magnitude, we have from Eq. (12) and (11)

$$\left.\frac{\partial J_{dom}}{\partial x}\right|_{x=L} = mq\delta(t - t_{tr})$$
$$\Rightarrow J_{dom}(L, t_{tr}) = mqL\delta(t - t_{tr}) \tag{13}$$

Finally, the total current density $\boldsymbol{J}_{total}(t)$ can be computed using the principle of superposition, as

$$J_{total}(x,t) = J_r(t) + J_{dom}(L, t_{tr}) \tag{14}$$

*2.3 Transit time*

The transit time is defined, as the time required reaching the domain from the cathode side to the anode side, so by definition:

$$t_{tr} \equiv L/v_{dom} \tag{15}$$

But the domain does not travel along the device length in a uniform velocity, rather $v_{dom} = v_{dom}(E_{dom}(x,t))$, hence we define

$$t_{tr} \equiv \sum_{\iota=1}^{N} t_\iota = \sum_{\iota=1}^{N} \frac{x_\iota}{v_{dom}(E_{dom}(x_\iota, t_\iota))} = \sum_{\iota=1}^{\xi-1} \frac{x_\iota}{v_{dom}(E_{dom}(x_\iota, t_\iota))} + \frac{\sum_{\iota=\xi}^{N} x_\iota}{v_{sat}} \tag{16}$$

where $v_{dom}$ is the domain velocity and the point $\xi$ has been defined by $v_{dom}(E_{dom}(\xi, \xi)) \equiv v_{sat}$. The device must be kept in its NDR region for proper operation, while the NDR region has been determined by the relation $dv_d/dE < 0$. So the applied electric field must be in such a magnitude that

$$E_{cr} < E_{appl} \leq E_{sat} \tag{17}$$

For computation, the carrier concentration $n(x,t)$ and domain electric field are regarded as constant in each of the cells and are evaluated at regular intervals. The cells are equally-spaced and made small enough to model the exact situation. The cells are also divided in time scale, using the relationship $t_\iota = x_\iota/v_{dom,\iota}$. The domain velocity at each cell has been evaluated by using the equal-area rule. The initial value of concentration and electric field has been taken from the contact of the cathode side, as domain formation probability is stronger there [1] and then they are updated in each cell according to the mathematical formulation described above. The end value of domain electric field has been set by the boundary conditions. The boundary conditions are that the device contacts are ohmic and the total current density must be continuous throughout the device. Zero electric field boundary conditions at the outer metal contact edges of the *n+* region (anode and cathode end) have been considered. The limits on the dimensions of space width (Δ*x*) and time width (Δ*t*) were derived from the physical consideration that the electron density should not change too much between neighboring space cells, *i.e.,* Δ*x* must be less than the Debye length [15]. Also, the integral of the electric field across the device must be equal to the applied voltage ($V_{DC}$). Mathematically, the boundary condition can be given by

$$E_{dom,\zeta} x_\zeta + E_r \sum_{\zeta \neq \kappa}^{N} x_\kappa = V_{DC}, \quad \zeta \epsilon \mathbb{N} \tag{18}$$

where $E_{dom,\zeta}$ can be given by the Maxwell's equation (2) and $E_r$ is the field outside the domain vicinity.

*2.4 Domain width*
The width of the domain is given by [1]

$$d = \frac{\varepsilon_0 \varepsilon_s}{q n_0}(E_{dom} - E_r) \tag{19}$$

## 3. Material properties and fitting parameters used
We have used the analytical expression proposed by Farahmand et al [9] to fit [25] with the recent reports by Bertazzi et al [7, 12]. All the parameters used in this paper to evaluate the velocity-field characteristics have been listed in Table 1 for GaN, and InN.

**Table 1.** Fitting and material parameters used in Eq (7) for approximate the steady state drift velocity-field characteristics of III-Nitrides. The parameters are valid for an applied electric field strength less than 800 kV cm$^{-1}$, an operating temperature of 300 K and doping concentration of $N_D = 10^{17}$ cm$^{-3}$.

| Parameters | GaN | | InN | |
|---|---|---|---|---|
| | Γ-M | Γ-A | Γ-M | Γ-A |
| $\mu_o$ (cm$^2$V$^{-1}$s$^{-1}$) | 1250 | 1000 | 3150 | 3150 |
| $E_{cr}$ (kVcm$^{-1}$) | 215 | 200 | 90 | 90 |
| $\upsilon_{sat}$(10$^7$cms$^{-1}$) | 1.0 | 1.3 | 1.00 | 1.00 |
| a | 8.0 | 5.5 | 5.75 | 5.5 |
| **$\beta_1$** | 4.5 | 4.25 | 3.20 | 3.10 |
| **$\beta_2$** | 0.7 | 0.7 | 0.85 | 0.90 |

## 4. Simulation, results and discussion
All the simulations have been performed for 300 K temperature and doping concentration $10^{17}$ cm$^{-3}$. The bias voltage 150 V, 90 V, and 30 V has been taken for 5 μm, 3 μm, and 1 μm device lengths for GaN, respectively and 18 V for 1 μm active length for InN, so that a domain electric field of 300 kV/cm is set up in the bulk for GaN and 180 kV/cm for InN. The saturation has been assumed at 800 kV/cm. By mathematical formulations presented, the domain velocity, electric field outside the domain vicinity, three dimensional evolution of domain dynamics, and the current density have been shown in figure 1, figure 2, figure 3, and figure 4 for GaN and InN-based Gunn diode.

From figure 1, it is evident that, the domain velocity decreases with the increasing domain field for GaN and InN, respectively. Figure 2 shows the analogous decrease in the field outside the domain vicinity for GaN and InN, respectively. The dynamic behavior of domain field increases with space and time until it reaches the saturation point, after which it has been assumed to be unchanged, shown in figure 3 for both materials. The three-dimensional evolution of the domain has been simulated for three active lengths: 5 μm, 3 μm, and 1 μm, while for InN, it has only shown for 1 μm active length. The current density due to domain field and the field outside the domain region with respect to time has been shown in figure 4 for GaN and InN, respectively, which also exhibits the frequency response for the materials in the respective active lengths. For GaN and InN the transit-time frequency has been found about 210 GHz and 224 GHz with current densities about 655kA/cm$^2$ and 691 kA/cm$^2$, respectively in 1 μm active length. The domain width increases almost in a linear pattern for both materials. For fitting purpose, our fitting parameter Δn has been adjusted to $10^{13}$. All the simulations have been performed by using MATLAB 7.8.0.347 (R2009a).

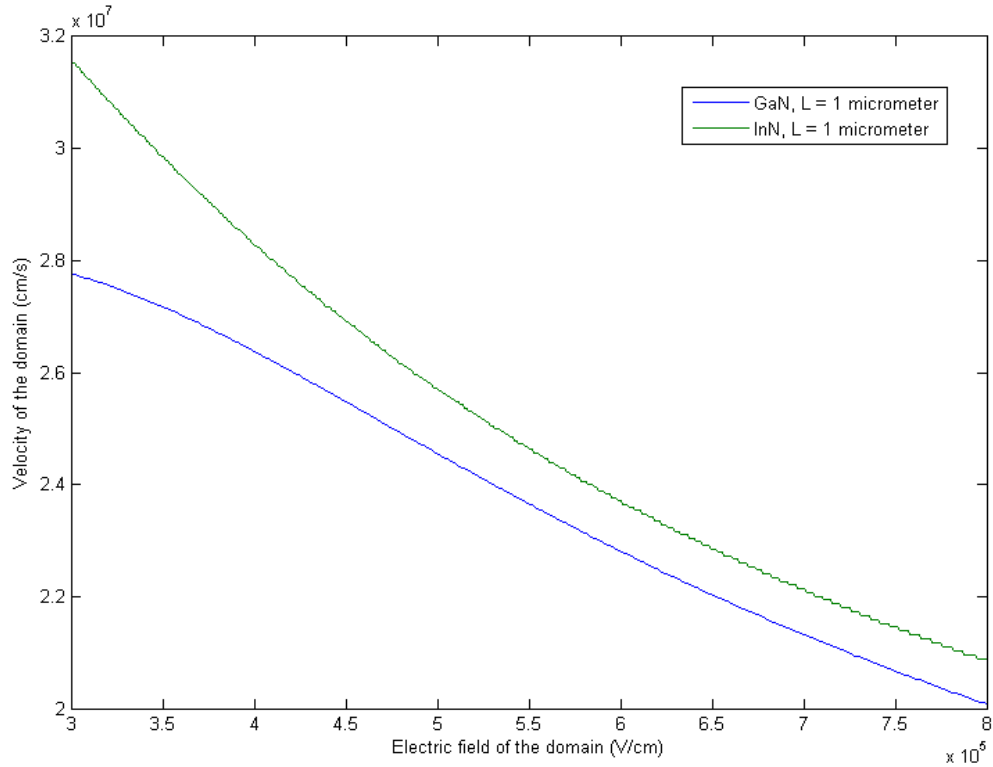

**Figure 1.** Variation of the domain velocity with domain electric field, determined by the equal-area rule for GaN and InN.

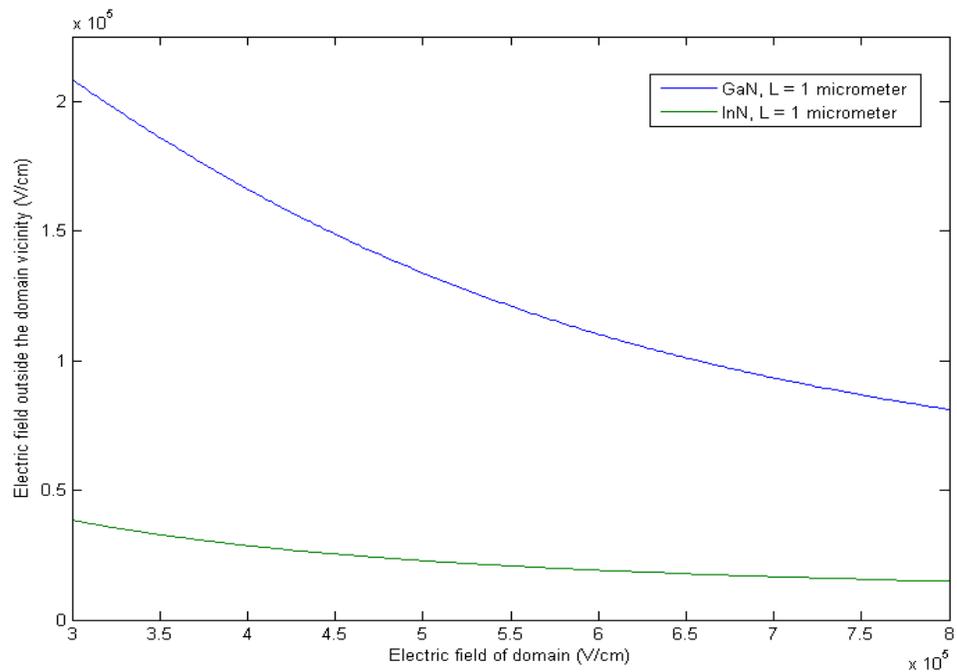

**Figure 2.** Variation of the field outside the domain vicinity with the domain field for GaN and InN.

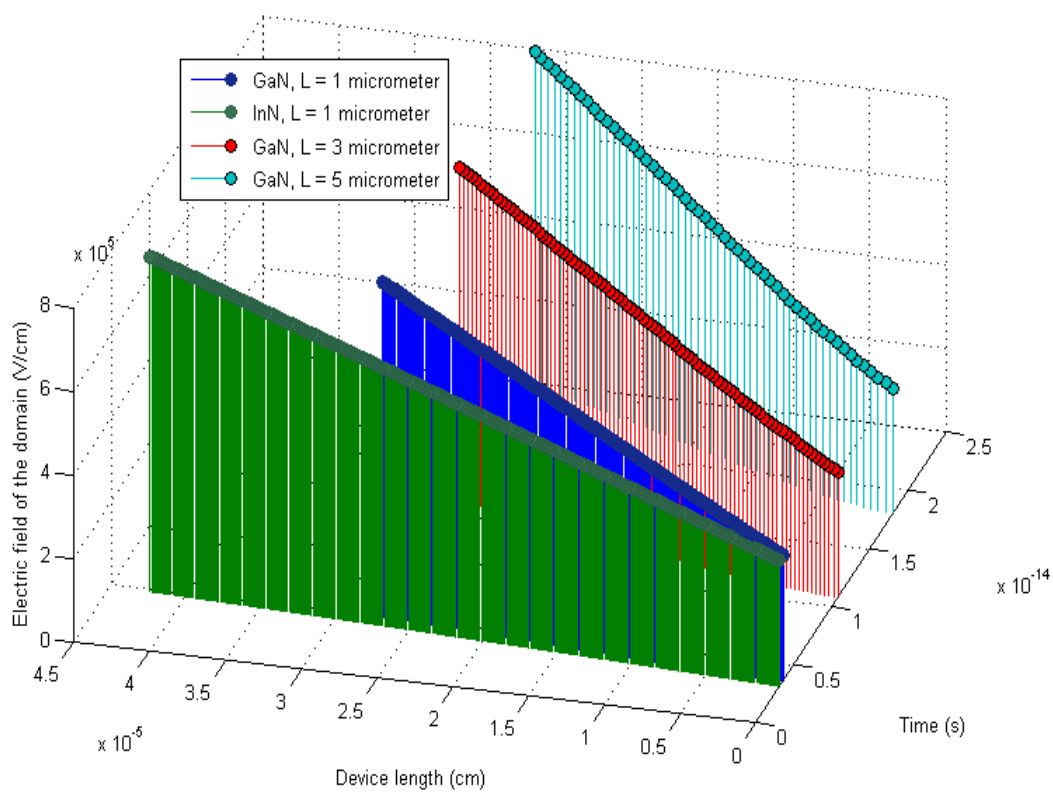

**Figure 3.** Evolution of the domain field with respect to space and time for GaN and InN of three different active lengths found by numerical solution of Maxwell's equation.

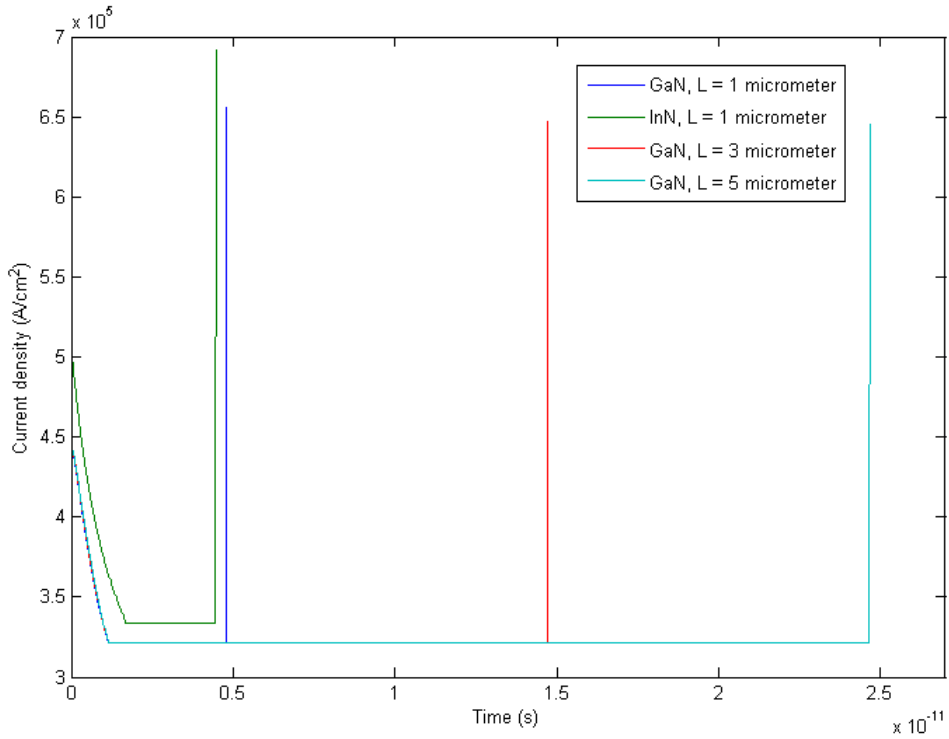

**Figure 4.** The current density with respect to time for a single cycle of operation for GaN and InN for different active lengths.

The perturbated region has been specified quantitatively unlike Roblin and Rohdin [24]. The domain width is linearly related to the domain electric field [1] and the domain electric field is governed by the carrier concentration according to the Maxwell's equation. Hence, to predict the domain width for the first time, it is necessary to have some constant factor. Our developed fitting parameter preciously does this which is absent in the model in Roblin and Rohdin [24]. Also, they have considered constant domain velocity for a particular applied bias, while we have used the varying domain velocity, as shown in figure 1, which is more realistic. Figure 2 exhibits the higher decay rate of the field outside of the domain region for InN than GaN due to the higher rate of domain field formation in InN than GaN in the same active length, as shown in figure 3.

There is no experimental report on GaN-based Gunn diodes, so far. However, some simulation and theoretical data are available. Alekseev and Pavlidis [18] found 87 GHz frequency response by using Medici in 3 μm active length. Panda *et al* [15] theoretically reported operating frequencies in between 100-120 GHz with DC biasing variation in an active length of 5 μm diode and 0.8-1.8 THz for 0.2 μm active length. To compare, we have evaluated the frequency response using our formulation and we have found around 41 GHz and 68 GHz in 5 μm and 3 μm active lengths at biasing voltage 150 V and 90 V, respectively. Our calculation exhibits about 17% discrepancy with Alekseev and Pavlidis [18], provided some of the material parameters used by them and us were different due the updated evaluation of those properties. They have used constant dielectric relaxation time and the classical velocity-field characteristic in their simulation, which is another reason for such difference and definitely seems unjustified at present. In the model given by Panda *et al* [15], there is no mention of analytical form of the velocity-field characteristics for GaN. Provided, the classical high-field velocity model employed for GaAs cannot be used to fit the measured velocity data for III-nitrides due to their dual-slope behavior [9, 12]. Also, there is no distinction between domain contribution and lower field contribution in their model, while in the current density curve, the oscillations have been reported. They have reported frequency for 0.2 μm active length which violates the Kroemer criterion. So the reported frequencies at 0.2 μm active length found by non-ballistic approach are highly questionable. The numerical result of Panda *et al* [15] contradicts that with both Alekseev and Pavlidis [17-19], and with the present work on frequency response and current density.

We claim that, this is the first report on InN-based Gunn diode. For InN a current density about 691kA/cm$^2$ is obtained at an active length of 1 μm with operating frequency about 224 GHz. Our simulated results indicate that InN dominates over GaN in relative small scale device, while GaN works better for longer devices. The early velocity overshoot effect of InN than GaN and larger saturation velocity of GaN than InN is the physical origin of these results.

**5. Conclusion**
Variation of domain velocity and field outside the domain with the field of the domain have been simulated and studied for the GaN and InN-based Gunn diode. The 3D evolution of domain with space and time, and domain width has also been studied. Finally, the frequency response has been computed. This is the first theoretical work for InN-based Gunn diode.